\documentclass[sigconf]{acmart}

\usepackage{multirow}
\usepackage{array}
\usepackage{latexsym}
\usepackage{booktabs}
\usepackage{amsmath}
\usepackage{pifont}
\usepackage{listings}
\usepackage{balance}

\usepackage{fancybox}   

\newenvironment{promptbox}
  {\begin{center}\setlength{\fboxsep}{10pt}
   \begin{Sbox}\begin{minipage}{0.94\linewidth}\small}
  {\end{minipage}\end{Sbox}\fbox{\TheSbox}\end{center}}

\lstdefinestyle{jsonplain}{
  basicstyle=\ttfamily\small,
  numbers=none,
  frame=single,
  rulecolor=\color{black},
  backgroundcolor=\color{gray!05},
  columns=fullflexible,
  keepspaces=true,
  showstringspaces=false,
  breaklines=true
}

\usepackage{subcaption} 

\usepackage[T1]{fontenc}

\usepackage[utf8]{inputenc}
\usepackage{xspace}

\usepackage{microtype}

\usepackage{graphicx}

\newcommand{\perf}{\textit{Perfect-Order}\xspace}
\newcommand{\worst}{\textit{Worst-Order}\xspace}
\newcommand{\random}{\textit{Random-Order}\xspace}
\newcommand{\modperf}{\textit{Moderately Perfect-Order}\xspace}
\newcommand{\inverse}{\textit{Moderately Inverted-Order}\xspace}

\usepackage[utf8]{inputenc}

\definecolor{deepgreen}{rgb}{0.0, 0.5, 0.0}
\definecolor{ForestGreen}{rgb}{0.0, 0.5, 0.0}

\colorlet{rqbg}{gray!30}           

\newcommand{\rqboxc}[1]{%
  \par\smallskip
  {\setlength{\fboxsep}{4pt}
   \noindent\fcolorbox{rqbg}{rqbg}{%
     \parbox{\dimexpr\linewidth-2\fboxsep\relax}{#1}%
   }}%
  \par\smallskip
}

\newcommand{\phead}[1]{\noindent {\bf #1}}
\newcommand{\uhead}[1]{\textit{#1}}

\AtBeginDocument{%
  }

\copyrightyear{2026}
\acmYear{2026}
\setcopyright{rightsretained}
\acmConference[ICSE '26]{2026 IEEE/ACM 48th International Conference on Software Engineering}{April 12--18, 2026}{Rio de Janeiro, Brazil}
\acmBooktitle{2026 IEEE/ACM 48th International Conference on Software Engineering (ICSE '26), April 12--18, 2026, Rio de Janeiro, Brazil}
\acmDOI{10.1145/3744916.3764524}
\acmISBN{979-8-4007-2025-3/26/04}

\begin{document}

\title{Order Matters! An Empirical Study on Large Language Models' Input Order Bias in Software Fault Localization}

\author{Md Nakhla Rafi}
\orcid{0009-0005-4707-8985}
\affiliation{%
  \institution{Software Performance, Analysis, \\and Reliability (SPEAR) Lab\\Concordia University}
  \city{Montreal}
  \country{Canada}
}
\email{mdnakhla.rafi@mail.concordia.ca}

\author{Dong Jae Kim}
\orcid{0000-0002-3181-0001}
\affiliation{%
  \institution{DePaul University}
  \city{Chicago}
  \country{USA}
}
\email{dkim121@depaul.edu}

\author{Tse-Hsun (Peter) Chen}
\orcid{0000-0003-4027-0905}
\affiliation{%
  \institution{Software Performance, Analysis, \\and Reliability (SPEAR) Lab\\Concordia University}
  \city{Montreal}
  \country{Canada}
}
\email{peterc@encs.concordia.ca}

\author{Shaowei Wang}
\orcid{0000-0003-3823-1771}
\affiliation{%
  \institution{University of Manitoba}
  \city{Winnipeg}
  \country{Canada}
}
\email{Shaowei.Wang@umanitoba.ca}

\renewcommand{\shortauthors}{Md Nakhla Rafi, Dong Jae Kim, Tse-Hsun (Peter) Chen, Shaowei Wang}

\begin{abstract}
Large Language Models (LLMs) show great promise in software engineering tasks like Fault Localization (FL) and Automatic Program Repair (APR). This study investigates the impact of input order and context size on LLM performance in FL, a crucial step for many downstream software engineering tasks. We test different orders for methods using Kendall Tau distances, including "perfect" (where ground truths come first) and "worst" (where ground truths come last), using two benchmarks that consist of both Java and Python projects. Our results indicate a significant bias in order; Top-1 FL accuracy in Java projects drops from 57\% to 20\%, while in Python projects, it decreases from 38\% to approximately 3\% when we reverse the code order. Breaking down inputs into smaller contexts helps reduce this bias, narrowing the performance gap in FL from 22\% and 6\% to just 1\% on both benchmarks. 
We then investigated whether the bias in order was caused by data leakage by renaming the method names with more meaningful alternatives. Our findings indicated that the trend remained consistent, suggesting that the bias was not due to data leakage.
We also look at ordering methods based on traditional FL techniques and metrics. Ordering using \textit{DepGraph}'s ranking achieves 48\% Top-1 accuracy, which is better than more straightforward ordering approaches like \textit{CallGraph\textsubscript{DFS}}. These findings underscore the importance of how we structure inputs, manage contexts, and choose ordering methods to improve LLM performance in FL and other software engineering tasks.
\end{abstract}




\maketitle

\section{Introduction}

Software development has significantly transformed with the emergence of Large Language Models (LLMs) like ChatGPT~\cite{chatGPT}. These tools have revolutionized how developers code, debug, and maintain software systems~\cite{zhang2023survey}. LLMs are widely adopted for their ability to simplify and accelerate development workflows, providing insights into complex tasks such as code generation and comprehension~\cite{abedu2024llm, lin2024llm}. 

Recent research has explored the use of LLMs in various software engineering tasks, including Fault Localization (FL)~\cite{autofl, testfreefaultlocalization} and Automatic Program Repair (APR)~\cite{AutoCodeRover, xia2024agentless}, which show great potential for automatically resolving real-world issues in large codebases. 
In particular, FL is a foundational step in the process, where the LLM processes structured lists to locate potential faulty code that requires fixing. Hence, high FL accuracy is instrumental to APR and automatic issue resolution.

While LLMs have demonstrated strong reasoning capabilities, prior research from other domains highlights a sensitivity to the order of input information. Studies have shown that LLMs perform better when information is presented in a sequence aligned with logical steps, with accuracy dropping significantly when the order is randomized~\cite{chen2024premise}. Additionally, LLMs exhibit a primacy effect, often prioritizing earlier information in prompts~\cite{wang2023primacy}. Although there are related studies on mathematical reasoning tasks, it is unclear whether such sensitivities extend to software engineering scenarios, such as FL. Since FL involves analyzing ordered lists of methods or elements, the sequence in which this information is presented may influence the model’s ability to identify faults.

This paper investigates how input order and context size affect the performance of large language models (LLMs) in Fault Localization (FL). We used the Defects4J~\cite{defects4j} and BugsInPy~\cite{widyasari2020bugsinpy} benchmarks, which are widely used datasets in software engineering for evaluating fault localization techniques. We conduct our experiments using two families of LLMs, ChatGPT 4o-mini and Deepseek-Chat. First, we assess whether the order of methods impacts the LLM’s ability to rank and identify faults by generating various input orders using Kendall Tau distance~\cite{cicirello2019kendall}, including \textit{perfect} (ground truth methods first) and \textit{worst} (ground truth methods last) orders. 
We found that the LLM’s performance is significantly influenced by input order. For instance, we observed that the Top-1 accuracy in the \textit{Defects4J} dataset decreased from 57\% to 20\%, and in the \textit{BugsInPy} dataset, it plummeted from 33\% to nearly 3\% when the list of methods was reversed. This suggests a strong bias towards the order of inputs, and the findings are consistent across the two studied LLMs. 

Next, we explore segmenting large inputs into smaller contexts using divide-and-conquer to address observed order biases. We observed that segmenting input sequences into smaller contexts reduces this bias; for example, the Top-1 gap between \perf and \worst rankings reduces from around 22\% in \textit{Defects4J} and 6\% in \textit{BugsInPy} at segment size 50 to just 1\% at segment size 10. 
We examined whether order bias was due to data leakage by renaming methods in the benchmarks with meaningful alternatives. The performance drop before and after renaming was minimal (Top-1 decreased by 9\% for segment size 50 and 4\% for size 10). However, the order bias between \perf and \worst remained, confirming that the bias stems from prompt ordering rather than data leakage.
Finally, we tested traditional FL and metrics-based ordering methods. We found that using FL techniques improved results, with \textit{DepGraph} outperforming \textit{Ochiai} by 16\% in Top-1 accuracy, while simpler strategies like \textit{CallGraph} and \textit{LOC} produced similar outcomes. These results highlight the importance of input order, context size, and effective ordering methods for enhancing LLM-based fault localization.

In summary, our contributions are as follows:
\begin{itemize}
    \item Method order significantly impacts LLM performance, as demonstrated by Top-1 accuracy in \textit{Defects4J}, dropping from 57\% in \perf (ground truths first) to 20\% in \worst (ground truths last). Similarly, for \textit{BugsInPy}, accuracy at Top-1 declines from 33\% to nearly 3\%.
    \item We demonstrate that dividing input sequences into smaller segments effectively mitigates order bias, the Top-1 gap between \perf and \worst rankings decreases from 22\% in \textit{Defects4J} and 6\% in \textit{BugsInPy} at segment size 50 to just 1\% at segment size 10.
    \item We find that order bias is due to prompt ordering, not data leakage, as renaming methods had minimal impact (Top-1 drop: 9\% for size 50, 4\% for size 10), while the bias between \perf and \worst remained.
    \item Ordering with different metrics and FL strategies significantly impacts outcomes. Ordering based on \textit{DepGraph} achieves 48\% Top-1 accuracy, 13.4\% higher than \textit{CallGraph\textsubscript{BFS}}. However, simpler methods like \textit{CallGraph\textsubscript{DFS}} reach 70.1\% Top-10 accuracy, highlighting their practicality in resource-constrained environments.

\end{itemize}

Our findings highlight the critical role of input order and segmentation in optimizing LLM-based fault localization. We believe our findings can also inspire future software engineering research that involves processing inputs as a list of software artifacts in the era of LLM.

\section{Background and Related Work} 
In this section, we first present the background on fault localization and large-language models. Then, we discuss related works. 

\subsection{Fault Localization}  
Fault Localization (FL)~\cite{wong2016survey} is a critical software engineering task that identifies specific program parts responsible for a failure. It is particularly essential in large and complex codebases, where manually finding faults can be time-consuming and error-prone. FL saves significant developer effort and serves as a cornerstone for many downstream software engineering tasks such as Automatic Program Repair (APR)~\cite{le2021automatic}, debugging automation~\cite{zamfir2010execution}, and performance optimization~\cite{woodside2007future}. The process begins with some indication of a fault, typically indicated by a failing test, which serves as the starting point. The input for FL often consists of a set of methods or code elements executed during the failing test case. FL aims to produce a ranked list of the most likely fault locations, providing developers with a focused starting point for investigation and resolution. Its significance lies in facilitating effective debugging and establishing the groundwork for workflows that automate and optimize the software development lifecycle.

\subsection{Related Work}
\phead{Spectrum-based and Supervised Fault Localization.} Traditional methods such as Spectrum-Based Fault Localization (SBFL) use statistical techniques to assess the suspiciousness of individual code elements.~\cite{abreu2007accuracy}. The intuition is that the code elements covered by more failing tests and fewer passing tests are more suspicious. While lightweight and straightforward, these techniques, such as Ochiai~\cite{abreu2009spectrum}, often struggle with achieving high accuracy in complex systems. 
To improve accuracy, supervised techniques like DeepFL~\cite{li2019deepfl} and Grace~\cite{lou2021boosting} incorporate features such as code complexity, historical fault data, and structural relationships using machine learning and Graph Neural Networks (GNNs). DepGraph~\cite{depgraph} further refines this approach by leveraging code dependencies and change history for improved fault ranking. 

\phead{LLM-Based Fault Localization.} 
Recent advances in Large Language Models (LLMs) have demonstrated significant potential for FL by leveraging their ability to analyze both code and natural language. Trained on extensive programming datasets, LLMs can understand code structure, interpret test failures, and even suggest fixes~\cite{autofl, wu2023large, pu2023summarization}. Building on these capabilities, LLM agents extend LLM functionalities by incorporating features like memory management~\cite{zhou2023recurrentgpt} and tool integration~\cite{roy2024exploring}, enabling them to execute tasks described in natural language autonomously. These agents can also adopt specialized roles, such as developers or testers, to enhance their domain-specific reasoning and improve problem-solving workflows~\cite{hong2024metagpt, white2024chatgpt}.

Several recent works have leveraged LLMs for FL. Wu et al.\cite{wu2023large} leverage test failure data to identify faulty methods or classes, enabling context-aware reasoning directly from the input. AutoFL\cite{autofl} enhances LLM capabilities by integrating tools to fetch and analyze covered classes and methods, providing additional insights for FL. AgentFL~\cite{qin2024agentfl} takes a more specialized approach, using agents with a Document-Guided Search method to navigate codebases, locate faults, and prioritize suspicious methods. In contrast, Agentless~\cite{xia2024agentless} simplifies FL with a three-phase workflow: localization, repair, and validation, eliminating the need for agents or complex tools. These tasks often involve handling large contexts, as LLMs process extensive lists of methods or code snippets, making the structure of input data a critical factor.

\subsection{Order Bias in LLM}
Prior research has demonstrated that Large Language Models (LLMs) exhibit sensitivity to input order, significantly influencing their reasoning and decision-making processes. Studies have shown that LLMs achieve higher accuracy when premises are structured sequentially according to logical reasoning steps, whereas randomized orderings lead to substantial performance drops~\cite{wang2023primacy, chen2024premise}. This phenomenon is particularly evident in deductive reasoning tasks, where presenting premises in the same order as a ground-truth proof enhances model accuracy, while shuffling them can result in a performance drop of over 30\%. Similarly, in mathematical problem-solving, LLMs struggle when problem descriptions are reordered, underscoring the importance of presenting input in a structured manner.

Despite these findings in other domains, the impact of input order on LLM-driven software engineering (SE) is still an open research question. To the best of our knowledge, this study is the first to investigate the impact of input ordering on LLM-based fault localization. Since fault localization involves analyzing ordered lists of methods, understanding the input order bias is critical for further improving LLM-based fault localization. Moreover, since many software engineering tasks involve analyzing a list of inputs, our findings may inspire future SE research and open a new research avenue.

\section{Methodology and Experiment Design}
\label{studyrq}
\begin{figure*}[t]
    \centering
    \includegraphics[width=\textwidth]{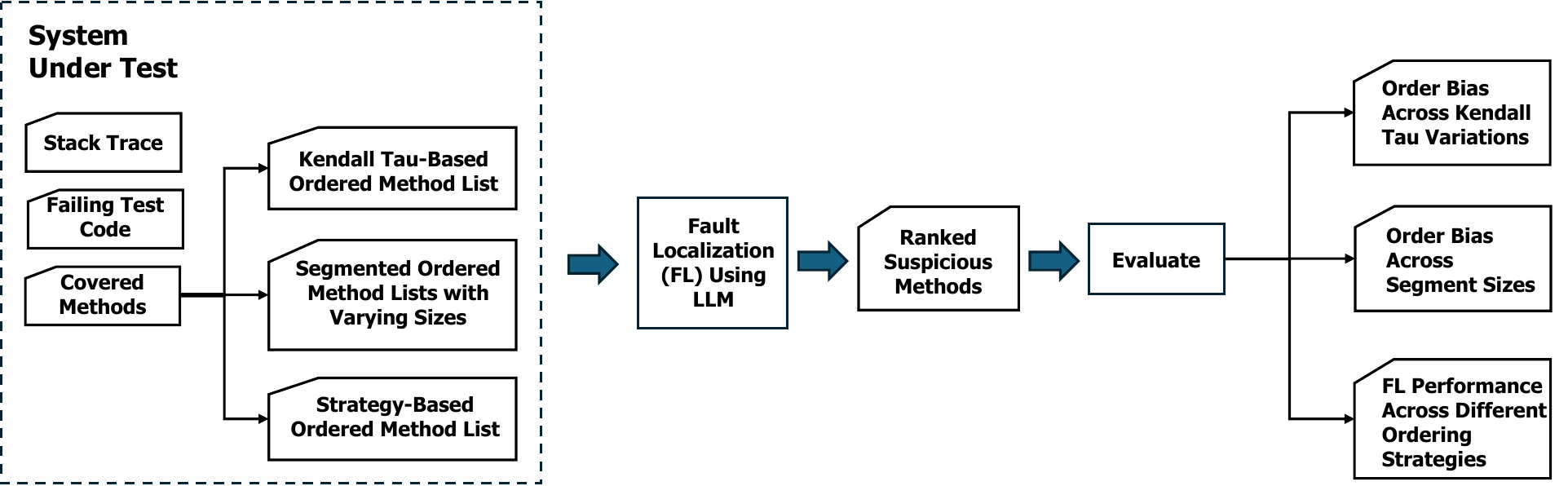} 
    \Description{An overview of our overall approach.}
    \caption{An overview of our overall approach.}
    \label{fig:overview}
\end{figure*}

This section describes the design of our experiments to study the impact of input ordering on fault localization.  Figure~\ref{fig:overview} shows the overall process. First, we collect code coverage information, including details about failing tests, stack traces, and the methods covered. Next, we generate various method orderings using the Kendall Tau distance as inputs for LLMs for fault localization. Finally, we evaluate the model's bias by calculating the Top-K accuracy across different orderings. Below, we discuss in more detail.

\begin{table}
\caption{Overview of the studied projects from Defects4J and BugsInPy.}
\centering
\scalebox{0.8}{
\setlength{\tabcolsep}{1pt} 
\begin{tabular}{l|l|rrrr}
    \toprule
    \textbf{Benchmark} & \textbf{Project} & \textbf{\#Faults} & \textbf{LoC} & \textbf{\#Tests} & \textbf{\#Fault-triggering Tests} \\
    \midrule
    \multirow{14}{*}{\textbf{Defects4J}}  
    & Cli          & 39  & 4K    & 94    & 66  \\
    & Codec        & 18  & 7K    & 206   & 43  \\
    & Collections  & 4   & 65K   & 1,286 & 4   \\
    & Compress     & 47  & 9K    & 73    & 72  \\
    & Csv          & 16  & 2K    & 54    & 24  \\
    & Gson         & 18  & 14K   & 720   & 34  \\
    & JacksonCore  & 26  & 22K   & 206   & 53  \\
    & JacksonXml   & 6   & 9K    & 138   & 12  \\
    & Jsoup        & 93  & 8K    & 139   & 144 \\
    & Lang         & 64  & 22K   & 2,291 & 121 \\
    & Math         & 106 & 85K   & 4,378 & 176 \\
    & Mockito      & 38  & 11K   & 1,379 & 118 \\
    & Time         & 26  & 28K   & 4,041 & 74  \\
    \cmidrule{2-6}
    & \textbf{Total} & \textbf{501} & \textbf{490K} & \textbf{15,302} & \textbf{901} \\
    \midrule
    \multirow{12}{*}{\textbf{BugsInPy}}  
    & cookiecutter  & 4  & 4.7K  & 300  & 5  \\
    & PySnooper         & 3  & 4.3K  & 73   & 2  \\
    & sanic        & 5  & 14.1K & 643  & 3  \\
    & httpie      & 5  & 5.6K  & 309  & 5  \\
    & keras         & 45  & 48.2K & 841  & 42 \\
    & thefuck             & 32  & 11.4K & 1,741 & 38 \\
    & black               & 15  & 96K   & 142  & 23 \\
    & scrapy           & 40  & 30.7K & 2,381 & 55 \\
    & luigi           & 33  & 41.5K & 1,718 & 36 \\
    & fastapi        & 16  & 25.3K & 842  & 31 \\
    & tqdm               & 9   & 4.8K  & 88   & 9  \\
    \cmidrule{2-6}
    & \textbf{Total} & \textbf{207} & \textbf{286.6K} & \textbf{8,078} & \textbf{249} \\
    \bottomrule
\end{tabular}}
\label{tab:merged_projects}
\end{table}

\subsection{Methodology}
\subsubsection{Prompt Design.} 
We use LLMs to rank the most suspicious methods in fault localization tasks by analyzing \textit{failing tests}, \textit{stack traces}, and \textit{covered methods}. We designed the prompts to be simple so we could better study the order bias. Figure~\ref{fig:prompt_example} shows an example of our prompt. It consists of two primary components: a 1) \textit{System Message} and a 2) \textit{User Message}, to guide the LLM in ranking suspicious methods. The \textit{system message} establishes the task by instructing the LLM to analyze a \textit{failing test}, its \textit{stack trace}, and a list of \textit{covered methods} during execution. The LLM ranks the top ten methods in descending order based on its analysis of suspicion. To ensure consistency in the generated output, the \textit{system message} specifies the required output format as a JSON structure, which includes method identifiers and their corresponding ranks. 

The \textit{user message} provides the input data specific to a \textit{failing test}, including the failing test code, the minimized \textit{stack trace}, and the \textit{covered methods}. Following prior works~\cite{autofl, wu2023large}, we retain only the information directly relevant to fault localization for \textit{stack traces}, discarding unrelated lines such as those from external libraries or other modules. This reduction enhances clarity and ensures that the LLM only processes essential data to identify the root cause of the failure. Covered methods are presented as an ordered list, serving as the candidate set for ranking. 

\begin{figure}
  \centering

  \begin{promptbox}
    \textbf{System Message:}\\
    You are a software debugging assistant. Your task is to analyze a system under test and rank the most suspicious methods.\\[4pt]
    \textbf{User Message:}\\
    Analyze the provided failing test, stack trace, and covered methods to localize faults and rank the top 10 most suspicious methods.\\[6pt]
    \textbf{Test Code:} 
    \texttt{\{test\_code\}}\\[4pt]
    \textbf{Stack Trace:} 
    \texttt{\{stack\_trace\}}\\[4pt]
    \textbf{Covered Methods:} 
    \texttt{\{covered\_methods\}}\\[8pt]
    The output should follow the JSON format below:\\[2pt]
    \textbf{JSON Format:}
    \begin{verbatim}
    {
      "methodB": "rank",
      "methodA": "rank",
      "..."
    }
    \end{verbatim}
  \end{promptbox}

  \Description{Prompt for Fault Localization.}
  \caption{Prompt for Fault Localization: (Top) system message; (Bottom) user message with test details and ranking request.}
  \label{fig:prompt_example}
\end{figure}

  
    
    




    

\subsubsection{Computing Kendall Tau Distance}

Kendall Tau (\(\tau\)) is a statistical measure used to quantify the similarity between two ranked lists by comparing the number of \textbf{concordant} and \textbf{discordant} pairs~\cite{cicirello2019kendall}. It provides a value between \(-1\) and \(1\), where \(1.0\) indicates a perfect match, \(0.0\) represents a random ranking, and \(-1.0\) signifies a fully reversed order.

To compute Kendall Tau, we use the formula:

\[
\tau = \frac{C - D}{C + D},
\]

\noindent where \(C\) is the number of \textbf{concordant pairs}, meaning two methods appear in the same order as in the reference ranking. \(D\) is the number of \textbf{discordant pairs}, meaning two methods appear in reverse order compared to the reference ranking. \(C + D\) is the total number of pairs, computed as:

\[
\frac{n(n-1)}{2},
\]
\noindent for \(n\) methods in the ranking.

\phead{Defining the Perfect Ranking.}  
The perfect ranking is derived from the execution sequence shown in the test case and method call order (see the top part of Figure~\ref{fig:method_call_order_kendall_tau}). Each method is assigned a unique \texttt{method\_id} based on its appearance in the \texttt{covered\_methods} list. The faulty ground truth method is always placed first (\texttt{method\_id: 1}), followed by methods in their exact execution order from the failing test. This ensures the ranking reflects the natural call graph order. From the code example in Figure~\ref{fig:method_call_order_kendall_tau}, the perfect ranking is \([1,2,3,4,5]\), where each method appears in the order it is executed. This serves as a baseline for comparison with other rankings.

\phead{Computing Kendall Tau for Different Rankings.}  
To analyze how deviations from the perfect ranking affect fault localization, we generate alternative rankings by shuffling the method order. The Kendall Tau score is computed for each by counting concordant and discordant pairs.
A concordant pair \((i, j)\) is one where \(i < j\) and their order matches the perfect ranking. For example, consider Figure~\ref{fig:method_call_order_kendall_tau}, in the perfect ranking \([1,2,3,4,5]\), the pairs \((1,2)\) and \((2,3)\) are concordant since they maintain the correct order. In contrast, a discordant pair \((i, j)\) is one where \(i < j\), but their order is reversed compared to the perfect ranking. In the worst ranking \([5,4,3,2,1]\), the pairs \((5,4)\) and \((4,3)\) are discordant since they appear in the opposite order.

As shown in Figure~\ref{fig:method_call_order_kendall_tau}, the perfect ranking \([1,2,3,4,5]\) results in \(\tau = 1.0\) since all pairs are concordant. A moderately shuffled ranking such as \([1,3,2,5,4]\) introduces some discordant pairs, reducing \(\tau\) to \(0.5\). A random ranking such as \([3,1,5,2,4]\) balances concordant and discordant pairs, yielding \(\tau = 0.0\). Finally, the worst ranking \([5,4,3,2,1]\) fully reverses the method order, resulting in \(\tau = -1.0\). By computing Kendall Tau for these variations, we assess how ranking order impacts fault localization accuracy.

\begin{figure*}
\centering
\footnotesize
\begin{lstlisting}%[caption={Example test case and method call order.}, label={lst:test_case}]
 // Test Case
 void testMethod1() {
     Example obj = new Example();
     String result = obj.method1("Start");
     assertEquals("Start -> Step2 -> Step3 -> Step4 -> Step5", result);
 }

 // Ground Truth Faulty Method
 String method1(String input) { return input + " -> " + method2(); }

 // Call Graph Methods
 String method2() { return "Step2 -> " + method3(); }
 String method3() { return "Step3 -> " + method4(); }
 String method4() { return "Step4 -> " + method5(); }
 String method5() { return "Step5"; }
\end{lstlisting}

\vspace{0.5cm} 

\resizebox{0.8\textwidth}{!}{
\begin{tabular}{l|l|c|c|c}
\toprule
\textbf{Ranking Name} & \textbf{Ranking} & \textbf{Concordant Pairs (C)} & \textbf{Discordant Pairs (D)} & \textbf{Kendall Tau} \\
\midrule
Perfect Order & [1, 2, 3, 4, 5] & 10 & 0 & 1.0 \\
Moderately Perfect & [1, 3, 2, 5, 4] & 7 & 3 & 0.5 \\
Random Order & [3, 1, 5, 2, 4] & 5 & 5 & 0.0 \\
Worst Order & [5, 4, 3, 2, 1] & 0 & 10 & -1.0 \\
\bottomrule
\end{tabular}
}
\Description{kendall tau table.}
\caption{(Top) The example test case and method call order illustrate how methods are executed in sequence, starting from the ground truth faulty method (\texttt{method1}) and progressing through the call graph (\texttt{method2} to \texttt{method5}). This order defines the \textbf{perfect ranking} \([1,2,3,4,5]\), where each method appears in the expected execution sequence. (Bottom) The table presents different method rankings evaluated using \textit{Kendall Tau}, a metric that measures ranking similarity. \textbf{Concordant pairs (C)} retain the correct relative order from the perfect ranking, while \textbf{discordant pairs (D)} appear in reverse order. The Kendall Tau score is calculated as \( \tau = \frac{C - D}{C + D} \), where \(1.0\) represents a perfect match, \(0.0\) indicates a random order, and \(-1.0\) signifies a fully reversed ranking.}

\label{fig:method_call_order_kendall_tau}
\end{figure*}

\subsection{Experiment Design}

\phead{Benchmark Dataset.} We experimented on 501 faults across 13 projects from the Defects4J benchmark (V2.0.0)~\citep{defects4j} and 207 faults from 11 projects from the BugsInPy benchmark~\cite{widyasari2020bugsinpy}. Both Defects4J and BugsInPy are widely used benchmarks in the software engineering community for fault localization~\cite{lou2021boosting, sohn2017fluccs, chen2022useful, zhang2017boosting,depgraph, autofl, rezaalipour2024fauxpy}. They provide a controlled environment for reproducing real-world bugs from a variety of projects, which differ in type and size. 
These benchmarks include both faulty and fixed project versions, along with associated test cases (including failing ones), metadata, and automation scripts, which facilitate research in FL, testing, and program repair. 

Table \ref{tab:merged_projects} gives detailed information on the projects and faults we use in our study. We excluded several projects from Defects4J and BugsInPy due to compilation errors, and the entirety of the coverage did not fit into one prompt because of the models' maximum input token size, which limited test coverage for most bugs. In total, we studied 708 faults with over 750 fault-triggering tests (i.e., failing tests that cover the fault).
Note that since a fault may have multiple fault-triggering tests, there are more fault-triggering tests than faults. 

\phead{Evaluation Metrics. }
We perform our fault localization process at the method level in keeping with prior work \citep{li2019deepfl, lou2021boosting, vancsics2021call, depgraph, autofl}. Namely, we aim to identify the source code methods that cause the fault. We apply the following commonly used metrics for evaluation:

\uhead{Accuracy at Top-N}. The Top-N metric measures the number of faults with at least one faulty program element (in this paper, methods) ranked in the top N. The results are a ranked list based on the suspiciousness score. Prior research \cite{parnin2011automated} indicates that developers typically only scrutinize a limited number of top-ranked faulty elements. Therefore, our study focuses on Top-N, where N is set to 1, 3, 5, and 10. 

Following prior LLM-based FL studies~\cite{autofl, wu2023large}, we did not use metrics like Mean First Rank (MFR) and Mean Average Rank (MAR) to measure how early faulty methods are ranked and their average position~\cite{lou2021boosting, li2019deepfl}. These metrics are unsuitable for LLM-based approaches because LLMs are inherently language models that do not generate explicit numeric scores.

\phead{Implementation and Environment.} 
To collect test coverage data and compute results for baseline techniques, we utilized Gzoltar~\cite{campos2012gzoltar}, an automated tool that executes tests and gathers coverage information. For the LLM-based components, we employed OpenAI's GPT-4o mini and DeepSeek's deepseek-chat. GPT-4o mini, which currently points to gpt-4o-mini-2024-07-18, has a context window of 128,000 tokens and can output 16,384 tokens at once~\cite{openai_gpt4o_2024}. Deepseek-Chat, which points presently to DeepSeek-V3, has a context window of 64,000 input tokens and a maximum of 8,000 output tokens~\cite{deepseek_api}. 
We used LangChain v0.2 to streamline the process of our experiment~\cite{langchain_docs_2024}. 
To minimize the variations in the output, we set the temperature parameter to 0.

\section{Experiment Results}
In this section, we present the results to our research questions. 

\subsection*{RQ1: Does the order in which the model processes code elements impact its performance?}
\label{section:rq1}
\begin{figure*}[htbp]
    \centering
    \begin{subfigure}[b]{0.48\linewidth}
        \centering
        \includegraphics[width=\linewidth]{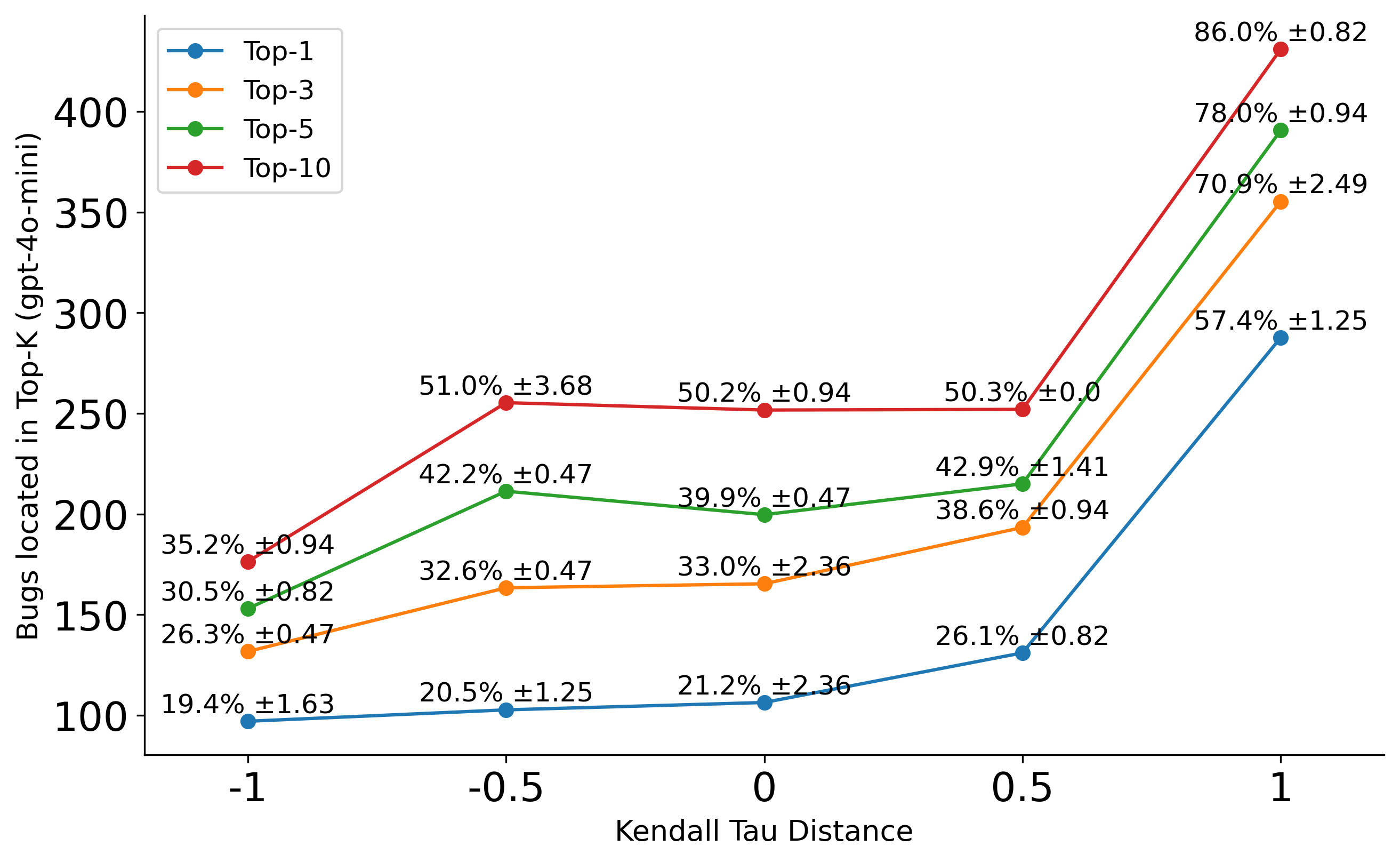}
        \caption{Defects4j - gpt-4o-mini}
    \end{subfigure}
    \hfill
    \begin{subfigure}[b]{0.48\linewidth}
        \centering
        \includegraphics[width=\linewidth]{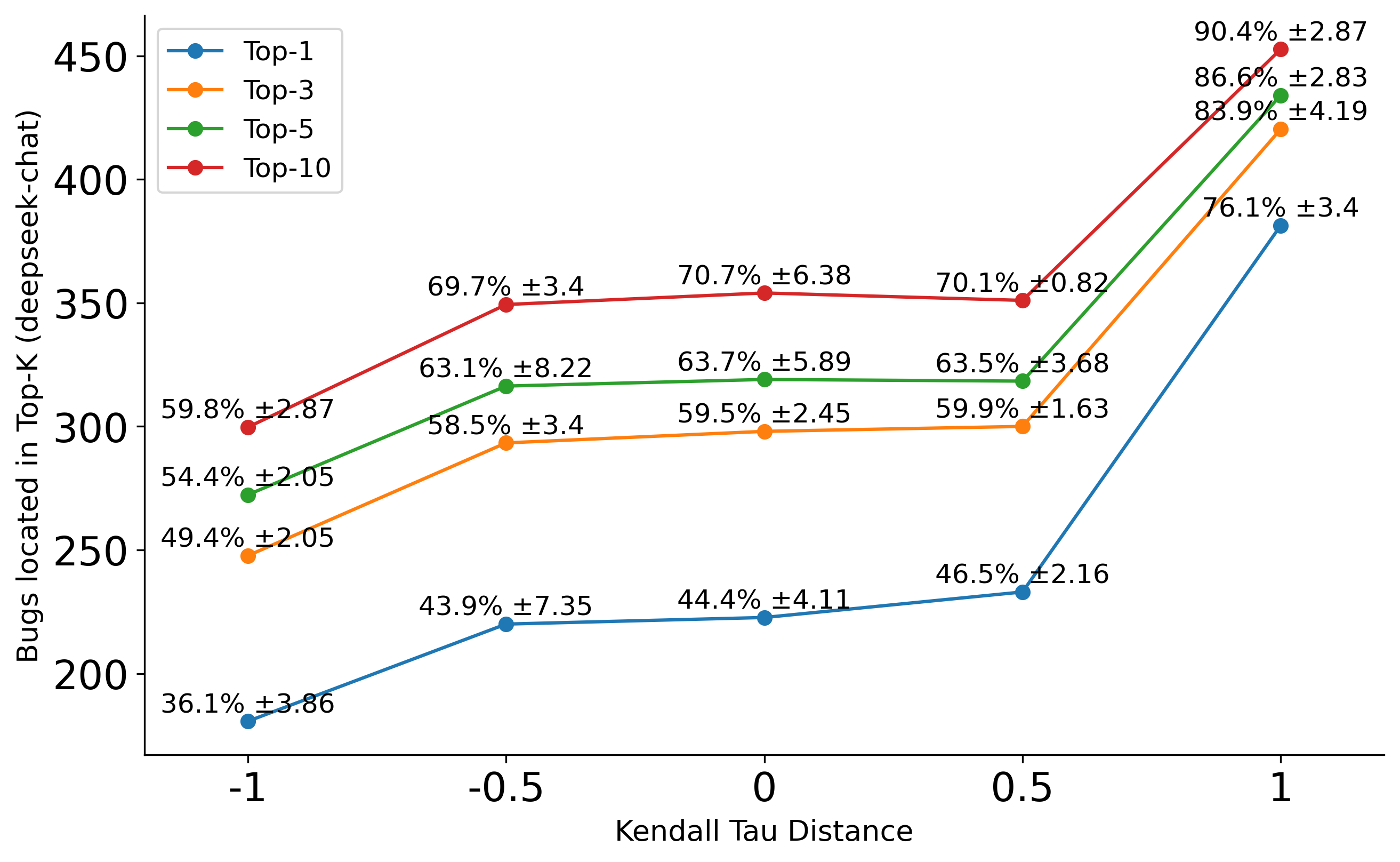}
        \caption{Defects4j - deepseek-chat}
    \end{subfigure}
    \begin{subfigure}[b]{0.48\linewidth}
        \centering
        \includegraphics[width=\linewidth]{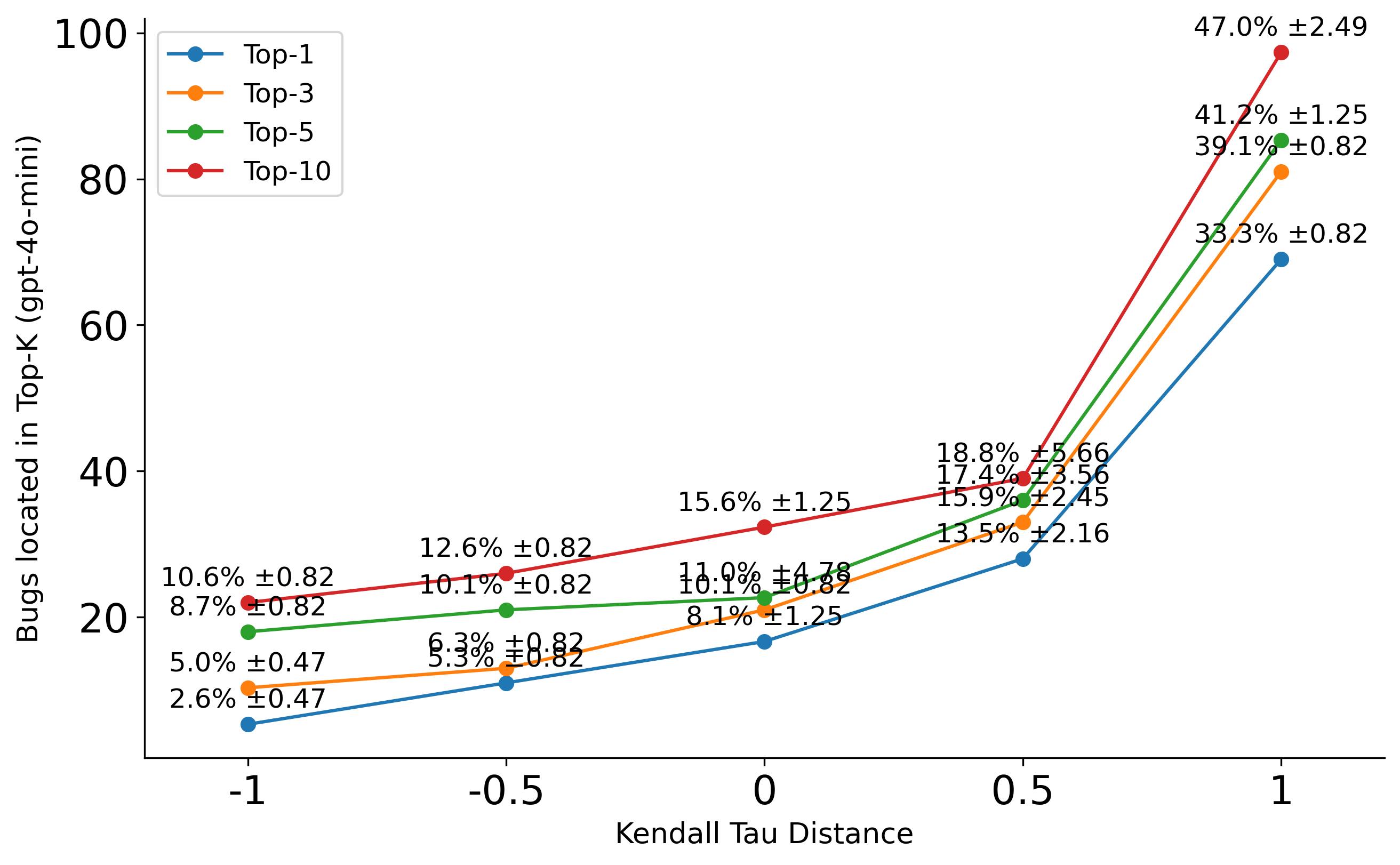}
        \caption{BugsInPy - gpt-4o-mini}
    \end{subfigure}
    \begin{subfigure}[b]{0.48\linewidth}
        \centering
        \includegraphics[width=\linewidth]{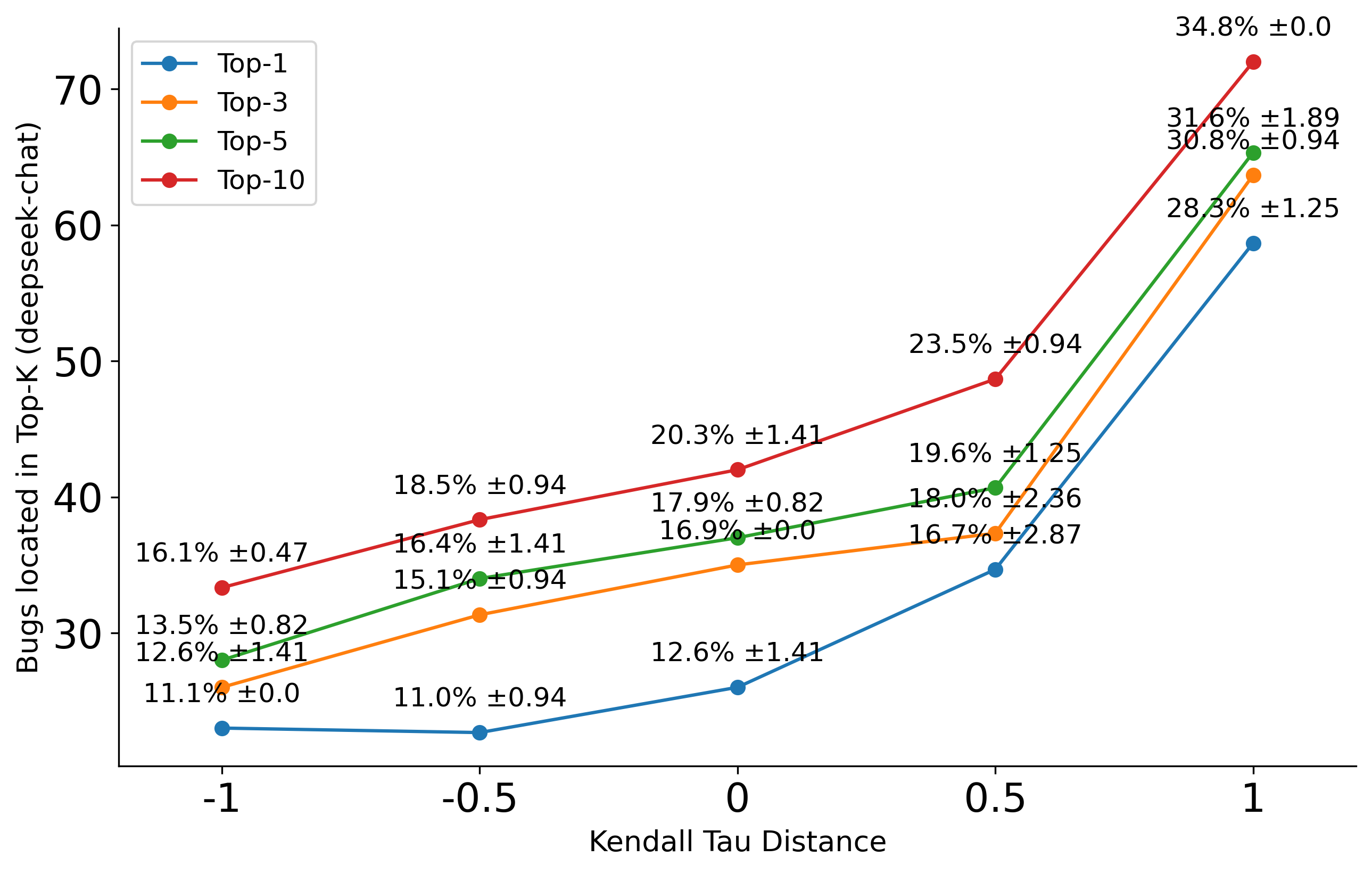}
        \caption{BugsInPy - deepseek-chat}
    \end{subfigure}
    \caption{The x-axis represents the number of bugs located, and the data points on the lines indicate the percentage of bugs identified out of the total (with standard deviation) at each Top-K position for various Kendall Tau (\(\tau\)) values.}
    \label{fig:rq_fig}
\end{figure*}

\phead{Motivation.} LLMs often struggle to reason over long input sequences, known as order bias, where the model prioritizes input tokens at the beginning or end of the sequence~\cite{wang2023primacy}. While order bias has been studied in NLP tasks, such as deductive and mathematical reasoning~\cite{chen2024premise}, its impact on software engineering tasks remains under-explored. Order is crucial in software engineering tasks, such as fault localization and program repair, where the model must reason over a long code sequence. Therefore, in this RQ, we investigate how code sequence order affects LLM accuracy in fault localization. 

\phead{Approach.} To study how code order sequences affect LLM-based fault localization, we create baselines with varying orderings for projects in \textit{Defects4J} (Java) and \textit{BugsInPy} (Python). The first baseline, \perf, places faulty methods (ground truths) at the top, followed by non-faulty methods, ordered by their \textit{call-graph} to minimize arbitrariness. Our intuition is that \perf serves as an idealized benchmark to test the hypothesis that prioritizing faulty methods should yield the highest accuracy if LLMs favor earlier orders due to their sequential processing nature. We then generate four additional baselines by adjusting the order using Kendall Tau distance~\cite{cicirello2019kendall}, which measures the correlation between two lists (i.e., 1 = perfect alignment with the \perf, -1 = complete misalignment with the \perf). From \perf, we derive: \textbf{\ding{192}} \random ($\tau = 0$; methods shuffled randomly), \textbf{\ding{193}} \worst ($\tau = -1$; faulty methods last), \textbf{\ding{194}} \modperf ($\tau = 0.5$; partial alignment), and \textbf{\ding{195}} \inverse ($\tau = -0.5$; partial misalignment). Comparing these baselines to \perf allows us to assess how deviations from the \perf affect FL results. Finally, we evaluate the model’s FL performance by ranking methods based on suspiciousness and measuring Top-K accuracy. For instance, a Top-1 score of 50\% indicates that 50\% of faulty methods were ranked first. We conduct experiments using \textit{gpt-4o-mini} and \textit{deepseek-chat} to determine whether the trends are generalizable.

\phead{Results.}
\textbf{\textit{LLMs exhibit a bias toward the initial input order, achieving higher accuracy at Top-1 for \perf compared to \worst across both models and benchmarks.}} Figure \ref{fig:rq_fig} shows the results of the experiments. For \perf, \textit{gpt-4o-mini} identifies 57.4\% of faults of \textit{Defects4J} in the Top-1 accuracy, while \modperf reduces the model's fault detection to 26.1\% ($\Delta$ 31.1\%). As Kendall Tau decreases, the accuracy declines further, reaching the lowest (19.4\%) for \worst, despite the \textbf{\textit{code context remaining identical, except for the code order}}. This trend was similar when detecting faults in \textit{BugsInPy} using \textit{gpt-4o-mini}, identifying approximately 33\% of faults in Top-1 using \perf, while the performance using \worst declined to only 2.6\%. 
The decline in Top-1 performance from \perf to \worst followed a similar pattern for the \textit{deepseek-chat} model across both benchmarks.
These results highlight key limitations in how LLMs process code, suggesting they may \textbf{\textit{rely more on surface-level patterns than on a deep understanding of code semantics}}. 

This trend persists across all other Top-K metrics. For Top-3, \textit{gpt-4o-mini} detects 70.9\% of faults in the \perf in \textit{Defects4J}, decreasing to 38.6\% for \modperf, which then stabilizes to 33\% for both \random and \inverse, then decreasing further to 26.3\% with the \worst. We see similar trends for Top-5 and Top-10, with \perf detecting the most faults, with 78\% and 86\% faults, respectively, compared to the lowest fault detection of 30.5\% and 35.2\% for \worst. Similar decrease in performance was seen when identifying faults in the Top-3, Top-5, and Top-10 categories in the Python-based benchmark \textit{BugsInPy}. Both models show a gradual decline in their performance as the order shifts from \perf to \worst as seen in Figure \ref{fig:rq_fig} (c) and (d).
\textbf{\textit{These findings suggest that LLMs are biased toward methods listed earlier in the input, indicating a potential order bias when analyzing code sequences.}} 

\textit{\textbf{The low variability in standard deviation (STDEV) across multiple runs suggests consistent order bias.}} To ensure the reliability of our findings on order bias, we conducted the experiments three times and reported the average. Across all Top-K results, the STDEV remains stable, ranging from 0.00 to 5.66 (i.e., methods) for \textit{gpt-4o-mini} and 0.0 to 8.22 for \textit{deepseek-chat}. For instance, the highest STDEV of 5.5 for Top-10 when using the \textit{gpt-4o-mini} indicates minimal variations, with only five methods changing position across runs. This consistency demonstrates that \textit{\textbf{order bias is not an artifact of randomness but a persistent limitation in how LLMs process code sequences.}} 

The prompt lengths range from 238 to 51,750 tokens, with the shortest 25\% of prompts having fewer than 1,876 tokens (on average, six methods) and the shortest 50\% having fewer than 4,734 tokens (on average, 20 methods). We did not observe much difference when the prompt sizes changed. In Defects4J, for the shortest 25\% of prompts (157 bugs), shifting the order from perfect to worst leads to a drop in Top-1 performance from 78.98\% to 43.95\%, marking a 35.03\% decline, which is similar to when using the full dataset (38\% decline). The trend is comparable in the 26–50\% prompt length bracket (129 bugs). Top-1 accuracy falls from 62.02\% to 15.50\%, a 46.52\% difference. \textbf{\textit{This shows that order biases exist regardless of the number of methods to be analyzed in the prompt, even when using smaller prompt sizes or fewer methods.}}

We further examined how random orderings affect stability using three independently shuffled random orders on Defects4J with temperature 0.3. The Top-1 results had a mean of 96.33 and a 95\% confidence interval of ±11.78. We observed similar variability in confidence interval across Top-3 (±10.21), Top-5 (±12.72), and Top-10 (±8.79). Compared to a fixed random order with temperature 0 (mean 106.33 ±5.56), the variance under random orderings is notably higher. \textit{\textbf{This confirms that random ordering introduces instability}}, and because a random ordering could, by chance, resemble a worst-case ordering, the performance drop seen in worst cases is not merely hypothetical, but plausible in practice.

\rqboxc{The LLM's fault localization performance is significantly impacted by input order. In Java projects, the Top-1 accuracy of both models decreases by nearly 40\% when the method list is reversed, suggesting a bias towards data presented earlier. A similar trend is observed in Python projects with different models.}

\subsection*{RQ2: Does limiting the context window help reduce the bias towards order?} 
\phead{Motivation.}
In RQ1, we identified order bias in the sequence in which methods are presented in the zero-shot prompt. 
We hypothesize that a larger context window might amplify the bias toward method order, as the LLM processes all methods simultaneously and may weigh their order more heavily when generating responses.
In this RQ, we investigate how the context window influences order bias. Specifically, we examine how segmenting the input sequence into smaller, independent segments affects the performance of LLMs in software engineering tasks, such as fault localization, where both context size and order play a crucial role in reasoning.

\begin{figure}[h]
  \centering

  \begin{promptbox}
    \textbf{Covered Methods (5 total):}\\
    \texttt{%
    1.~methodA (Ground truth)\\
    2.~MyClass (Constructor)\\
    3.~helper1\\
    4.~helper2\\
    5.~anotherHelper
    }\\[6pt]

    \textbf{Prompt for Segment 1}\\[2pt]
    \textbf{Covered Methods in This Segment:} \texttt{methodA, MyClass, helper1}\\[2pt]
    \textbf{Prompt:} Analyze the failing test, stack trace, and covered methods to rank the most suspicious methods.\\[2pt]
    \textbf{Output (JSON):}
\begin{verbatim}
{
  "methodA": 1,
  "helper1": 2,
  "MyClass": 3
}
\end{verbatim}

    \textbf{Prompt for Segment 2}\\[2pt]
    \textbf{Previous Rankings from Segment 1:} \texttt{methodA > helper1 > MyClass}\\
    \textbf{Remaining Covered Methods in This Segment:} \texttt{helper2, anotherHelper}\\[2pt]
    \textbf{Prompt:} Analyze the failing test, stack trace, and covered methods to rank the most suspicious methods. Incorporate historical rankings while analyzing newly covered methods. Update the rankings accordingly.\\[2pt]
    \textbf{Output (JSON):}
\begin{verbatim}
{
  "methodA": 1,
  "helper1": 2,
  "helper2": 3,
  "anotherHelper": 4,
  "MyClass": 5
}
\end{verbatim}
  \end{promptbox}

  \caption{Context segmentation for fault localization: the initial ranking is generated from a subset of covered methods; the next segment updates the ranking by incorporating previous results while analyzing newly covered methods.}
  \label{fig:context_segmentation_prompt}
\end{figure}

\phead{Approach.}
We investigate whether a divide-and-conquer approach can reduce this bias, where the input sequence is split into smaller contexts and each subset is reasoned individually. Figure \ref{fig:context_segmentation_prompt} illustrates a simplified example of how we divide the coverage and put them in the prompts. We divide an ordered list of \( N \) methods, \( M=\{m_1, m_2, \dots, m_N\} \), into \( \max(\lceil N / S \rceil, 1) \) contiguous segments. Each segment \( M_i \subseteq M \) (for \( i = \{1, 2, \dots, J\} \)) contains up to \( S \) methods, ensuring \( |M_i| \leq S \). If \( S > N \), the entire list \( M \) forms a single segment (\( J = 1 \)). For this study, we experiment with five segment sizes \( S \in \{10, 20, 30, 40, 50\} \). In each segment  \( M_i \), the model ranks the Top-K suspicious methods \( R_i \), and the results \( R_i \) are summarized into \( G_i \). For the subsequent segment \( S_{i+1} \), the prompt includes both the \( M_{i+1} \) and the contextual information from \( G_i \). This iterative context-passing approach allows the model to re-rank methods based on combined information from previous segments. We analyze how the context window impacts order bias in fault localization by incrementally varying the segment size (\(S\)). Specifically, we compare performance across two extreme ordering sequences: \perf (\(\tau = 1\)) and \worst (\(\tau = -1\)) (defined in RQ1) to assess whether the iterative context-passing effectively mitigates order bias, improving reliability across diverse ordering sequences. We run the experiments for projects in \textit{Defects4J} and \textit{BugsInPy} using the \textit{gpt-4o-mini} model.

\phead{Results.} \textbf{\textit{The size of the context window impacts fault localization results, with larger context windows exhibiting a stronger order bias.}} Table~\ref{tab:split_comparison_merged} presents the Top-K scores across different context segments for \perf and \worst. For \textit{Defects4J}, when the context is provided in larger segments (e.g., segment size 50), the model detects 278 bugs (55.5\%) in Top-1 with the \perf, while the \worst identifies only 170 bugs (33.9\%), around 22\% fewer bugs than the \perf. The large difference in Top-1 shows a significant order bias towards the order of the input method list. This is also evident in the Top 3, 5, and 10. For example, in the Top-10, the Perfect ranking reaches 408 (81.4\%) compared to 292 (58.3\%) for the Worst ranking, detecting around 23\% more bugs. A similar pattern is observed in \textit{BugsInPy}, where the model detects 46 bugs (22.2\%) in Top-1 with the \perf at segment size 50, while the \worst detects only 34 (16.4\%), reinforcing the impact of order bias.

\textbf{\textit{As the segment size decreases, the difference between the \perf and \worst becomes smaller across all Top-K.}} For \textit{Defects4J}, at a segment size of 40, the model detects approximately 54\% of bugs in Top-1 with the \perf, which is 20\% more than the 34\% bugs detected with the \worst ranking. This pattern holds for Top-3, Top-5, and Top-10 as well. At a segment size of 30, the difference in Top-1 narrows further to 17\%, with \perf identifying 51\% of bugs compared to 34\% for \worst. When the segment size is reduced to 20, the \perf detects around 49\% of bugs in Top-1, while \worst detects 37\%, shrinking the difference to 12\%. This trend continues for Top-3, Top-5, and Top-10, where the performance gap between the two rankings becomes progressively smaller. For \textit{BugsInPy}, at segment size 20, \perf achieves a Top-1 accuracy of 42 (20.3\%), while \worst reaches 37 (17.9\%), again showing a decreasing order bias as segment size reduces.

At the smallest segment size of 10, there is nearly no difference in Top-1 (only a 1\% gap between \perf and \worst) for \textit{Defects4J}. Interestingly, for Top-3, Top-5, and Top-10, the model performs slightly better using the \worst compared to the \perf. This trend is also evident in \textit{BugsInPy}, where \worst slightly outperforms \perf in Top-3, Top-5, and Top-10 at segment size 10. These findings suggest that \textbf{\textit{as segment sizes decrease, the order bias toward the input order diminishes, and this effect is consistent across both Defects4J and BugsInPy benchmarks.}} However, using smaller segments implies issuing multiple prompts to the model, which could increase both latency and computational cost in real-world settings. Our results suggest that order sensitivity is not limited to long inputs but also can appear in shorter lists, extending to other prompt structures (e.g., ranking files before methods). Moreover, while our study focuses on zero-shot prompting, broader prompting strategies such as few-shot examples, chain-of-thought reasoning, or dynamic retrieval could potentially mitigate this sensitivity and are worth exploring in future work.

\rqboxc{As the context window size decreases, the order bias significantly diminishes across both Defects4J and BugsInPy. The Top-1 gap between \perf and \worst rankings reduces from around 22\% (55.5\% vs. 33.9\%) in Defects4J and 6\% (22.2\% vs. 16.4\%) in BugsInPy at segment size 50 to just 1\% at segment size 10. Larger context windows tend to increase bias, whereas smaller context windows help reduce it.}

\begin{table}
    \centering
        \caption{Comparison of fault localization performance across \perf and \worst orders with varying segment sizes, merging both Defects4J and BugsInPy. Each row shows the number of bugs (and their corresponding percentages) located in the Top-1, Top-3, Top-5, and Top-10 ranks.}

    \resizebox{\columnwidth}{!}{
    \begin{tabular}{l|l|c|c|c|c|c}
    \toprule
    \textbf{Benchmark} & \textbf{Ordering} & \textbf{Seg. Size} & \textbf{Top-1} & \textbf{Top-3} & \textbf{Top-5} & \textbf{Top-10} \\
    \midrule
    \multirow{10}{*}{\textbf{Defects4J}}
    & Perfect & 10 & 217 (43.3\%) & 295 (58.9\%) & 313 (62.5\%) & 338 (67.5\%) \\
    & Worst   & 10 & 211 (42.1\%) & 298 (59.5\%) & 330 (65.9\%) & 362 (72.3\%) \\
    \cmidrule{2-7}
    & Perfect & 20 & 247 (49.3\%) & 311 (62.1\%) & 343 (68.5\%) & 374 (74.7\%) \\
    & Worst   & 20 & 186 (37.1\%) & 265 (52.9\%) & 288 (57.5\%) & 335 (66.9\%) \\
    \cmidrule{2-7}
    & Perfect & 30 & 261 (52.1\%) & 326 (65.1\%) & 347 (69.3\%) & 382 (76.2\%) \\
    & Worst   & 30 & 175 (34.9\%) & 236 (47.1\%) & 266 (53.1\%) & 309 (61.7\%) \\
    \cmidrule{2-7}
    & Perfect & 40 & 270 (53.9\%) & 328 (65.5\%) & 347 (69.3\%) & 390 (77.8\%) \\
    & Worst   & 40 & 171 (34.1\%) & 223 (44.5\%) & 249 (49.7\%) & 284 (56.7\%) \\
    \cmidrule{2-7}
    & Perfect & 50 & 278 (55.5\%) & 338 (67.5\%) & 368 (73.5\%) & 408 (81.4\%) \\
    & Worst   & 50 & 170 (33.9\%) & 224 (44.7\%) & 248 (49.5\%) & 292 (58.3\%) \\
    \midrule
    \multirow{10}{*}{\textbf{BugsInPy}}
    &   Perfect & 10 & 40 (19.3\%) & 45 (21.7\%) & 46 (22.2\%) & 49 (23.7\%) \\
     &   Worst & 10 & 41 (19.8\%) & 50 (24.2\%) & 52 (25.1\%) & 60 (29.0\%) \\
        \cmidrule{2-7}
     &   Perfect & 20 & 42 (20.3\%) & 52 (25.1\%) & 52 (25.1\%) & 54 (26.1\%) \\
     &   Worst & 20 & 37 (17.9\%) & 49 (23.7\%) & 52 (25.1\%) & 57 (27.5\%) \\
        \cmidrule{2-7}
     &   Perfect & 30 & 40 (19.3\%) & 50 (24.2\%) & 53 (25.6\%) & 55 (26.6\%) \\
     &   Worst & 30 & 37 (17.9\%) & 48 (23.2\%) & 51 (24.6\%) & 53 (25.6\%) \\
        \cmidrule{2-7}
     &   Perfect & 40 & 44 (21.3\%) & 50 (24.2\%) & 52 (25.1\%) & 54 (26.1\%) \\
     &   Worst & 40 & 36 (17.4\%) & 49 (23.7\%) & 51 (24.6\%) & 56 (27.1\%) \\
        \cmidrule{2-7}
     &   Perfect & 50 & 46 (22.2\%) & 56 (27.1\%) & 57 (27.5\%) & 61 (29.5\%) \\
     &   Worst & 50 & 34 (16.4\%) & 41 (19.8\%) & 46 (22.2\%) & 51 (24.6\%) \\
    \bottomrule
    \end{tabular}
    }
    \label{tab:split_comparison_merged}
\end{table}

\subsection*{RQ3: Is the order effect due to data leakage?}
\label{rq3}
\phead{Motivation.} 
In prior RQs, we found that the sequence in which methods are listed within zero-shot prompts can create bias. This raises concerns regarding potential data leakage within large language models, as the LLMs may have recognized the ground truth faulty methods when they appear early in the input. Here, we examine whether the observed order effect is a consequence of data leakage.

\phead{Approach.}
We study data leakage issues by renaming the variables and method names in the code. If data leakage issues are present, renaming the methods should give very different results~\cite{balloccu-etal-2024-leak}. 
First, we replace each method name for the systems under test in Defects4J with meaningful alternatives generated by an LLM model, specifically \textit{GPT-4o-mini}. We take these alternative method names and use a static code parsing tool to traverse the abstract syntax trees (ASTs) to rename all methods accordingly. We ensured no leakage between renaming and evaluation by performing renaming in separate API sessions and excluding those logs from localization prompts. Renamed identifiers were applied using a static AST-based tool, with no shared context across stages.

We then repeat the context segmentation experiment from RQ2, but only with Segment sizes 50 and 10 in this case, to check if the same trend holds, specifically. Namely, does the bias reduce when the context window size decreases? For the orderings, we use the perfect \( \tau = 1\) and worst \( \tau = -1\) order. To reduce experiment costs and runtime, we conducted fault localization experiments in this RQ using only the \textit{gpt-4o-mini} model on \textit{Defects4J}, as previous RQs showed that order bias remains consistent across different benchmarks and models.

\phead{Results.} \textbf{\textit{The results confirm that order bias persists with renamed methods, indicating that the effect is not due to data leakage.}} Table \ref{tab:rename_split} presents the Top-K scores across different context segments for \perf and \worst using the alternate renamed methods. When the context window is provided in larger segments (e.g., a segment size of 50) using the \perf (\( \tau = 1\)), the model detects 253 bugs, which accounts for 50.5\% in the Top-1 results. In contrast, with the \worst approach, the model identifies only 152 bugs, representing 30.3\%—around 20\% fewer than the \perf results. Even after changing the method names, there remains a significant difference in the Top-1 results, indicating the presence of an order bias. This bias is not solely due to the model memorizing its training data. The effect is also noticeable in the Top 3, Top 5, and Top 10 results.

\textbf{\textit{As the segment size decreases (for example, to 10), the difference between the performance of the model (\perf) and the worst-case scenario (\worst) diminishes across all Top-K values.}} With the smallest segment size, the model detects 208 bugs (41.5\%) using the \perf metric, while it detects 197 bugs (39.3\%) using the \worst metric. For the top-3, top-5, and top-10 cases, the model performs slightly better or achieves comparable results when using the \worst metric compared to the \perf metric. This further demonstrates that reducing the context size for large datasets can minimize order bias, and this bias is not a result of data leakage issues.

We also find that \textbf{\textit{there is no significant data leakage issue on the benchmark data, following prior studies}}~\cite{zhou2025lessleak, xu2024benchmark}. Our results show a modest performance drop after renaming: for segment size 10, the Top-1 score using \perf drops from 217 to 208 (a 4\% decrease), and for segment size 50, from 278 to 253 (a 9\% reduction). While these results suggest that some degree of memorization may exist, the overall impact on performance appears limited. This aligns with findings from LessLeak-Bench~\cite{zhou2025lessleak}, which reported minimal leakage in Defects4J (0.41\%), and recent work by Ramos et al.~\cite{ramos2024large}, which showed that newer models like LLaMa3.1 exhibit limited signs of leakage. Since our models (GPT-4o-mini and DeepSeek) are trained on comparable or larger datasets, we expect similar behavior. However, as highlighted in recent studies~\cite{zhou2025lessleak}, particularly on BugsInPy, where leakage is more pronounced (11\%), this remains an open concern in LLM research. Notably, we observe a trend that larger context windows may amplify potential leakage, suggesting that segment size can influence how much the model “recalls” training data.

\rqboxc{Even after renaming methods, order bias persisted, confirming that it is due to prompt ordering rather than data leakage. The FL performance difference before and after renaming was minimal, with Top-1 scores decreasing by 9\% (278 to 253) for segment size 50 and 4\% (217 to 208) for size 10, suggesting no significant data leakage issue.}

\begin{table}
    \centering
        \caption{A comparison of fault localization performance across techniques and segments after changing the method names to address the data leakage issue. The table shows Defects4J's bugs detected in the Top-1, Top-3, Top-5, and Top-10 positions using \perf and \worst across various segments.}

    \resizebox{\columnwidth}{!}{
    \begin{tabular}{l|c|c|c|c|c}
    \toprule
        \textbf{Ordering} & \textbf{Seg. Size} & \textbf{Top-1} & \textbf{Top-3} & \textbf{Top-5} & \textbf{Top-10} \\ 
    \midrule
        Perfect & 10 & 208 (41.5\%) & 260 (51.9\%) & 292 (58.3\%) & 322 (64.3\%) \\
        Worst & 10 & 197 (39.3\%) & 262 (52.3\%) & 295 (58.9\%) & 331 (66.1\%) \\
        \midrule
        Perfect & 50 & 253 (50.5\%) & 323 (64.5\%) & 347 (69.3\%) & 393 (78.4\%) \\
        Worst & 50 & 152 (30.3\%) & 205 (40.9\%) & 245 (48.9\%) & 290 (57.9\%) \\

    \bottomrule
    \end{tabular}
    }
    \label{tab:rename_split}
\end{table}

\subsection*{RQ4: How do different ordering strategies influence fault localization performance?}
\label{rq3:section}
\phead{Motivation.} We find that LLMs may have order biases toward \perf when investigating a list of methods for FL. However, in practice, such ground truth ordering is unknown. Hence, in this RQ, we investigate whether ordering methods based on the static or dynamic nature of the code or using existing FL techniques can help LLMs achieve better FL results.

\phead{Approach.} We explore four types of ordering: (1) \textit{Statistical-based} and (2) \textit{Learning-based}, which are directly derived from FL techniques, and (3) \textit{Metric-based} and (4) \textit{Structure-based}, which are grounded in static code features and not specifically tied to FL. The first two approaches leverage dynamic execution data or advanced models trained on FL tasks, making them more targeted for identifying faults. In contrast, the latter two approaches are agnostic to FL techniques. Hence, they may lack the specificity needed for accurately identifying faults, as they do not directly utilize FL data. By integrating ordering strategies with the rich contextual information in the prompt template (see Figure~\ref{fig:prompt_example}), including test code, stack traces, and coverage data, we aim to strengthen LLMs' reasoning about the most relevant areas of the program, ultimately improving fault localization. Similar to RQ3, we run FL experiments only on \textit{Defects4J} using \textit{gpt-4o-mini} to reduce running time and experimental cost, as previous findings showed that the observed trends remain consistent across different models and benchmarks.

For \textit{Metric-based} ordering, we use \textit{Lines of Code (LOC)}, ranking methods in descending order of their lines of code. Longer methods are often more complex and fault-prone~\cite{herraiz2010beyond}, making LOC a simple yet effective heuristic for prioritization. For \textit{Structure-based} ordering, we consider the structure of the call graph associated with each failing test. Specifically, we use \textit{Call Graph\textsubscript{DFS}}, which prioritizes deeper methods by traversing the call graph using depth-first search (DFS), and \textit{Call Graph\textsubscript{BFS}}, which highlights immediate dependencies by traversing the call graph using breadth-first search (BFS). By explicitly encoding dependency relationships, we investigate whether these structural insights can aid LLMs in reasoning about fault propagation within the program and enhance fault localization.

\textit{Statistical-based} ordering relies on dynamic execution data. For this, we use \textit{Ochiai}, which prioritizes methods most likely to contain faults, offering insights beyond static metrics or structural heuristics. \textit{Ochiai} is a lightweight unsupervised Spectrum-Based Fault Localization (SBFL) technique~\cite{sasaki2020sbfl} based on the intuition that methods covered by more failing tests and fewer passing tests are considered more suspicious (e.g., faulty). Its suspiciousness score is computed as:

{\small
\begin{equation*}
Ochiai(a_{ef}, a_{nf}, a_{ep}) = \frac{a_{ef}}{\sqrt{(a_{ef} + a_{nf}) \times (a_{ef} + a_{ep})}}
\end{equation*}
}

Here, \(a_{ef}\), \(a_{nf}\), and \(a_{ep}\) denote the number of failed and passed test cases that execute or do not execute a code statement. Scores range from 0 to 1, with higher values indicating higher fault likelihood. We order methods by aggregating their statement-level scores. 

Finally, for \textit{Learning-based} ordering, we use \textit{DepGraph}, which is the state-of-the-art supervised FL technique based on a graph neural network model~\cite{depgraph} that transforms the rich static and dynamic code information into a graph structure. It trains a graph neural network to rank faulty methods by analyzing structural code dependencies and code change history. 

\begin{table}
        \caption{Comparison of fault localization performance using different ordering strategies with the percentage of bugs found across 501 total faults from Defects4J.}

    \centering
    \resizebox{\columnwidth}{!}{
    \begin{tabular}{ll|c|c|c|c}
    \toprule
       \multicolumn{2}{l|}{ \textbf{Technique} }& \textbf{Top-1} & \textbf{Top-3} & \textbf{Top-5} & \textbf{Top-10} \\ 
       \midrule
        \multicolumn{2}{l|}{\textit{\underline{Learning-based}}} & & & & \\
         & {DepGraph} & 242.0 (48.3\%) & 338.0 (67.5\%) & 386.0 (77.0\%) & 419.0 (83.6\%) \\
        \midrule
        \multicolumn{2}{l|}{\textit{\underline{Structure-based}}} & & & & \\
        & {CallGraph\textsubscript{BFS}} & 175.0 (34.9\%) & 252.0 (50.3\%) & 294.0 (58.7\%) & 343.0 (68.5\%) \\
        & {CallGraph\textsubscript{DFS}} & 173.0 (34.5\%) & 253.0 (50.5\%) & 305.0 (60.9\%) & 351.0 (70.1\%) \\
        \midrule
        \multicolumn{2}{l|}{\textit{\underline{Statistical-based}}} & & & & \\
        & {Ochiai} & 164.0 (32.7\%) & 252.0 (50.3\%) & 293.0 (58.5\%) & 342.0 (68.3\%) \\
        \midrule
        \multicolumn{2}{l|}{\textit{\underline{Metric-based}}} & & & & \\
        & {LOC} & 163.0 (32.5\%) & 256.0 (51.1\%) & 289.0 (57.7\%) & 351.0 (70.1\%) \\

    \bottomrule
    \end{tabular}
    }
    \label{tab:technique_comparison}
\end{table}

\phead{Results.}
\textbf{\textit{The choice of ordering strategy is critical in LLM's ability to localize faults, with FL-derived ordering using \textit{DepGraph} detecting almost 13.4\% more faults in Top-1 compared to the next highest Top-1, achieved by \textit{Call Graph\textsubscript{BFS}}}}.  Table~\ref{tab:technique_comparison} highlights the model's effectiveness across different ordering strategies. \textit{DepGraph} identifies 13.4\%  more faults in the Top-1 rank. We see similar trends among other Top-K, where \textit{DepGraph} identifies 16\% more faults in Top-3, 16.1\% in Top-5, and 13.5\% in Top-10. This performance difference is expected, as \textit{DepGraph} excels at ranking faulty methods higher on the list through its advanced fault localization capabilities. The additional faults localized by \textit{DepGraph} across all Top-K ranks reinforce our earlier observation that \textbf{\textit{improved ordering strategies enable the model to prioritize the most suspicious methods earlier}}. 

Despite the importance of order bias, actual FL methods like \textit{DepGraph} provide significantly better fault localization than techniques that do not use LLMs. For instance, \textit{DepGraph}'s Top-K results are higher: 299 (Top-1), 382 (Top-3), 415 (Top-5), and 449 (Top-10) compared to the results from LLM-based methods. The results suggest that \textit{\textbf{while LLMs can help with tasks such as ranking faulty methods, domain-specific methods (like DepGraph) are still superior for accurate results}}. 

Ordering through \textit{Ochiai} reveals an interesting trend: \textbf{\textit{leveraging simple statistical metrics for ordering can enable LLMs to improve fault localization by better prioritizing fault-prone methods}}. In its standalone form, \textit{Ochiai} achieves a Top-1 performance of 101. However, when its ranking is provided to the LLM, performance rises to 164, showing that even a simple statistical method can be significantly enhanced through LLM reasoning. This finding indicates that while \textit{Ochiai}, a lightweight statistical approach, does not match the accuracy of ordering through \textit{DepGraph}, it can still effectively assist in fault localization, particularly when computational efficiency or simplicity is a priority.
\textit{Ochiai} offers LLMs a more straightforward way to rank methods based on test outcomes, which aligns well with their ability to process observable patterns. 

\textbf{\textit{Simple static-based ordering strategies can match or even outperform more complex FL-derived ordering across all Top-K ranks.}} For instance, \textit{CallGraph\textsubscript{BFS}}, which prioritizes methods closer to failing tests, identifies 175 bugs in Top-1 (34.9\%), slightly outperforming \textit{CallGraph\textsubscript{DFS}} with 173 bugs (34.5\%) and achieving a higher Top-1 accuracy (32.7\%) compared to the more complex \textit{Statistical-based} ordering. A similar trend is observed across the remaining Top-K ranks, where \textit{CallGraph\textsubscript{DFS}} either matches or slightly outperforms \textit{Statistical-based} ordering, with differences ranging from 0\% to 0.02\%. Additionally, when comparing \textit{Statistical-based} ordering with \textit{Metric-based} methods, it shows comparable performance in Top-1 ($\Delta 0.2\%$) and Top-3 ($\Delta 0.8\%$), but outperforms in Top-5 ($\Delta 0.8\%$) and Top-10 accuracy ($\Delta 1.8\%$). This suggests that static methods, which are computationally less demanding, can still be effective for fault localization. Hence, these findings emphasize the practicality of \textbf{\textit{simpler static-based methods as viable alternatives to more complex FL techniques}}.

\rqboxc{While ordering helps rank faults, LLMs struggle with complex relationships. Simpler static-based methods, like \textit{CallGraph\textsubscript{BFS}}, perform comparably to more complex \textit{Statistical-based} ordering like \textit{Ochiai} in fault localization. Our findings highlight the practicality of static-based methods as efficient alternatives to complex FL techniques, particularly in resource-constrained environments.}

\section{Discussion and Implications}

\subsection{Implications of Ordering Strategies}  
Our findings show that the order of inputs significantly impacts the performance of large language models (LLMs) in FL. This highlights the need for thoughtful ordering strategies. Metrics-based ordering, drawn from traditional techniques like \textit{DepGraph} and \textit{Ochiai}, prioritizes suspicious methods and improves accuracy. For instance, \textit{DepGraph} achieved the highest Top-1 accuracy, demonstrating the effectiveness of advanced strategies. In contrast, simpler methods like \textit{CallGraph} and \textit{LOC} performed well across a broader range of cases, making them suitable for resource-limited situations.

When clear ordering metrics are unavailable, randomizing input orders can serve as a fallback to minimize potential biases introduced by positional effects. Additionally, refining prompts to emphasize context rather than sequence and training LLMs on diverse input sequences could further reduce order bias and improve their robustness. These insights indicate that aligning ordering strategies with task requirements and model capabilities is essential for optimizing workflows in LLM-based FL.

\subsection{Effectiveness of Segment-Based Strategies.}
The segment-based approach reduces order bias by keeping the input size small, allowing the model to reason over information step by step in smaller contexts. Specifically, we find that a context size of 10 minimizes bias, leading to similar Top-K results for both the \perf and \worst cases, where both share the same code context. However, as the context window increases, order bias becomes more influential, affecting the LLM's ability to reason over long sequences of code. Future research could focus on identifying optimal segment sizes that adjust based on task complexity and the amount of available input.

\section{Threats To Validity}
\textit{Threats to internal validity} may arise from our technique implementations and experimental scripts. To address this, we have thoroughly reviewed and implemented our code using state-of-the-art frameworks like Langchain~\cite{langchain_docs_2024}. Additionally, we utilized widely adopted tools such as Gzoltar~\cite{campos2012gzoltar} for test coverage analysis and Java Call Graph~\cite{gousiosg2023javacallgraph} for extracting call relationships, ensuring robustness in our fault localization process.

\textit{Threats to external validity} may be tied to the benchmarks used. We mitigate this by evaluating our approach on \textit{Defects4J}~\cite{defects4j} and \textit{BugsInPy}~\cite{widyasari2020bugsinpy}, two widely used benchmarks containing diverse real-world faults in Java and Python projects. These datasets are commonly used in fault localization research, making our findings more applicable to broader software engineering tasks.

\textit{Threats to construct validity} may arise from data leakage, where LLMs might recognize faulty methods due to memorization rather than genuine reasoning. To address this, we renamed method names with meaningful alternatives and found that performance remained consistent, confirming that order bias is caused by prompt ordering rather than data leakage.

\section{Conclusion}
This work highlights several areas for future research. Order bias may influence the performance of large language models (LLMs) in tasks beyond FL, such as program repair, test case prioritization, and code refactoring. Investigating how order bias affects these tasks and whether similar solutions can be applied would be beneficial. Additionally, specific prompts incorporating domain knowledge, such as code semantics and dependency graphs, could enhance contextual understanding and reduce reliance on positional hints. Lastly, exploring new evaluation metrics considering the significance of input order and context size will help us better understand how LLMs operate in software engineering tasks. We have made all data and scripts related to this work publicly available~\cite{AnonymousSubmission}.

\bibliographystyle{ACM-Reference-Format}
\bibliography{references}

\end{document}